\documentclass[apj]{emulateapj}

\bibpunct{(}{)}{;}{a}{}{,}

\slugcomment{in press, ApJL}

\shorttitle{Forming early-type galaxies in groups}
\shortauthors{Kautsch et al.}

\begin{document}

\title{Forming Early-type Galaxies in Groups Prior to Cluster Assembly}

\author{Stefan J. Kautsch\altaffilmark{1}, Anthony
H. Gonzalez\altaffilmark{1}, Christian A. Soto\altaffilmark{1},
Kim-Vy H. Tran\altaffilmark{2}, Dennis Zaritsky\altaffilmark{3} 
and John Moustakas\altaffilmark{4}}

\altaffiltext{1}{Department of Astronomy, University of Florida, Gainesville, FL 32611-2055}
\altaffiltext{2}{Institute for Theoretical Physics, Universit\"at Z\"urich, CH-8057 Z\"urich, Switzerland}
\altaffiltext{3}{Steward Observatory, University of Arizona, Tucson, AZ 85721}
\altaffiltext{4}{Center for Cosmology and Particle Physics, New York University, 
4 Washington Place, New York, NY 10003}


\begin{abstract}

We study a unique proto-cluster of galaxies, the supergroup
SG1120-1202.  We quantify the degree to which morphological
transformation of cluster galaxies occurs prior to cluster assembly in
order to explain the observed early-type fractions in galaxy clusters
at $z$=0. SG1120-1202 at $z$$\sim$0.37 is comprised of four gravitationally
bound groups that are expected to coalesce into a single cluster by
$z$=0. Using {\it HST} ACS observations, we compare the morphological
fractions of the supergroup galaxies to those found in a range of
environments. We find that the morphological fractions of early-type
galaxies ($\sim$60\%) and the ratio of S0 to elliptical galaxies (0.5) 
in SG1120-1202 are very similar to clusters at
comparable redshift, consistent with pre-processing in
the group environment playing the dominant role in establishing the
observed early-type fraction in galaxy clusters.

\end{abstract}



%%%%%%%%%%%%%%%%%%%%%%%%%%%%%%%%%%%%%%%%%%%%%%%%%%%%%%%%%%%%%%%%%%%%%%%%%%%

\section{Introduction}

Our understanding of the role of environment in determining galaxy
properties remains incomplete primarily because of the complicated
correlations among galaxy properties and the difficulty in
establishing the correspondence between local galaxies and their
progenitors.  While many properties, including morphology and star
formation rate, vary strongly as a function of environment
\citep[e.g.,][]{dressler80,lewis02,gomez03} the relative importance
of different physical processes in driving these variations remains
controversial \citep{park07}. Here, we focus on identifying the
environment in which predominantly late-type field galaxies are
``transformed" into early-type cluster galaxies.

Two general classes of physical scenarios have been proposed as
mechanisms for driving morphological transformations: local processes
like mergers \citep[e.g.,][]{toomre77} or tidal interactions
\citep[e.g.,][]{mastropietro05} and global processes such as ram
pressure stripping \citep[e.g.,][]{gunn72}, evaporation and
strangulation \citep[e.g.,][]{larson80} and harassment
\citep[e.g.,][]{moore99}.  Local processes are most effective in
groups because of the low relative velocities \citep{barnes85} whereas
global processes are most effective in clusters where the gravitational
potential is deep but the relative velocities of the galaxies are
high.

A compelling empirical case has been made that pre-processing in the
group environment is the dominant mechanism driving the observed
early-type fractions in clusters
\citep{zabludoff98,kodama01,helsdon03}. However, numerical simulations
have recently suggested that intermediate mass clusters accrete their
galaxies as individual systems, and that therefore any transformation
must occur in the cluster environment \citep{berrier08}.  Evidently,
the key in resolving this question lies in unambiguously tagging and
studying the distant galaxies that lie in today's massive clusters.

The natural environment to begin such a study is that of groups at
intermediate redshift \citep[e.g.,][]{wilman05,mulchaey06}.  However,
not all groups will be accreted by clusters and there exists a wide
dispersion in group properties at intermediate redshifts
\citep{poggianti06}.  To avoid these pitfalls, we 
focus on a unique
system,\footnote{ Based on observations with the NASA/ESA Hubble Space
Telescope, obtained at the Space Telescope Science Institute, which is
operated by AURA, Inc., under NASA contract NAS 5-26555; and based on
data collected at the VLT (072.A-0367, 076.B-0362, and 078.B-0409),
which is operated by ESO and the Magellan Telescope, which is operated
by the Carnegie Observatories.} \objectname{SG1120-1202} (hereafter
SG1120).  SG1120 is a gravitationally bound structure consisting of four
X-ray luminous groups at $z$$\sim$0.37, whose dynamics indicate that it
will collapse to form a cluster of similar mass to Coma by $z$=0
\citep{gonzalez05}.  Therefore, the question is relatively unambiguous
here: do the galaxy morphologies in SG1120 already match those
seen in local clusters of comparable mass or will they need to be
transformed after the groups coalesce?  
While SG1120 represents only one unique evolutionary path, most clusters
continue to accrete groups at late times and the group properties observed
in SG1120 are broadly relevant for assessing the impact of this late-type accretion.
In this Letter we use the
standard cosmology ($\Omega_{M}$=0.3, $\Omega_{\Lambda}$=0.7, 
$H_{0}$=70~km~s$^{-1}$~Mpc$^{-1}$) unless stated otherwise.

%%%%%%%%%%%%%%%%%%%%%%%%%%%%%%%%%%%%%%%%%%%%%%%%%%%%%%%%%%%%%%%%%%%%%%%%%%%

\section{Observations}

We measure morphologies using {\it HST} ACS F814W imaging (Cycle 14)
composed of ten pointings that cover
18$\arcmin$.4$\times$11$\arcmin$.8, with single orbit depth at each
location. Spectra were obtained with the VLT (VIMOS \& FORS2)~(Feb.~2004 and
2007) and Magellan (LDSS3)~(Feb.~2006) and ground based imaging with the VLT (VIMOS; Feb.~2003 and 2006) 
and the KPNO Mayall (FLAMINGOS; Feb.~2006) telescopes. The spectroscopic target
selection was based on Vega magnitude-limited catalogs
($R$$\leq$22.5 or $K_s$$\leq$20). The observations yield redshifts for 364
galaxies, including 156 confirmed SG1120
members on the {\it HST} ACS mosaic. 
Confirmed members are defined to lie within the 2~$\sigma$ velocity limits 
defined by the lowest and highest redshift groups (890 km s$^{-1}$ and 1551 km s$^{-1}$, respectively), which in turn corresponds to 0.349$\leq$$z$$\leq$0.377.
The velocity dispersions for the groups are measured using the biweight estimator \citep[rostat;][]{beers90}.

For the subsequent computation we use $M^{\ast}_V$=-21.28 
(m$_{814}$=19.1, Vega)	 
as in \citet[][]{postman05}. For the photometric filter
conversions including evolution correction we use the formulae from
\citet{fabricant00} which have a 0.1 mag associated uncertainty.

%%%%%%%%%%%%%%%%%%%%%%%%%%%%%%%%%%%%%%%%%%%%%%%%%%%%%%%%%%%%%%%%%%%%%%%%%%%

\section{Morphological Classification}
\subsection{Quantitative classification}

We use GIM2D \citep{simard02,simard08} for the quantitative
morphological classification.  GIM2D performs automated bulge/disk
decomposition and measures the object's asymmetry.  We compare our
morphological results with the reference data set of \citet{simard08}
who used the same quantitative classification on a morphological study
of high-redshift galaxy clusters and groups in the ESO Distant Cluster
Survey.

The galaxies are modeled using an $r^{1/4}$ profile for the bulge and
an exponential profile for the disk, and the models are convolved with the
point spread function (PSF) generated by TinyTim \citep{krist93}. For
each object, we calculate PSFs at the locations of the members in the
raw ACS images, and then drizzle \citep{fruchter02} them together to 
generate a composite PSF
for each object.

For each galaxy, we use the GIM2D bulge-to-total ratio (BT) and image
smoothness index (S2) \citep{simard08} inside two half light radii; S2
measures the overall smoothness of the galaxy with respect to the
model.

\subsubsection{Reliability tests with simulations} \label{simul}

A reliability check of GIM2D and error quantification is performed by
inserting artificial galaxies (created with GALIMAGE/GIM2D) 
in the real ACS frames that are then
analyzed in the same way as the observed galaxies. We follow the
description given by \citet[][]{simard02,simard08}.  The simulated
galaxies have random BT and inclination values between 0 and 1 
and 0\degr and 85\degr, respectively, and cover the magnitude
range of our targets.

We derive the averaged standard deviation for 
BT (=0.042) and S2 (=0.016) with the aid of the simulations. 
We then perform bootstrap resampling to 
compute a fractional error of 4\% using the standard deviations determined the 
simulations. Therefore we assume an uncertainty of $\pm$4\%  for the
quantitative morphological fractions derived throughout this letter.

\subsection{Visual Classification}

To complement our quantitative morphologies, five of our team
members\footnote{SJK, AHG, CAS, KVHT, DZ}
visually classified all confirmed cluster members according to
a simplified Hubble scheme: ellipticals (E), lenticulars (S0), spirals
(S), and irregulars (I). For this analysis the irregular
classification includes interacting and low surface brightness
galaxies as well as dwarfs. To standardize the classifications, we use
a training set based upon the sample from \citet{fabricant00}. For the
subsequent analysis, we adopt the most common assigned morphological
type when the classification from the individuals differs.

\subsection{Visual vs. Quantitative Classification}

\begin{figure}
\plotone{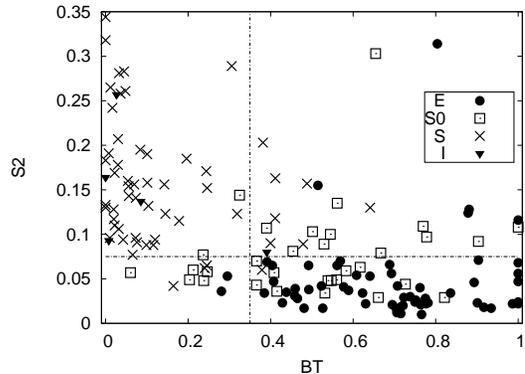}
\caption{Comparison of the visual morphologies (as indicated in the legend) 
to the GIM2D values. The quantitative bulge-to-total ratio (BT) is plotted on the $x$ axis
and smoothness index (S2) on the $y$ axis. 
The horizontal and vertical line indicates
the values separating our morphological classes based upon the
criteria by \citet{simard08}.}\label{fig1}
\end{figure}

Figure~\ref{fig1} shows the distribution of the visual
types (E, S0, S, and I) and their GIM2D values.  A correlation is
evident in Fig.~\ref{fig1}, with E+S0 galaxies
typically having 0.4$\le$BT$\le$1 (right) and S galaxies BT$<$0.4
(left). Asymmetric structures,
measured with S2, increase in late-type galaxies as expected. 
However, the divisions are not sharp and each class
contains some galaxies that lie outside these regions.
Asymmetric structures
such as rings, spiral arms, HII regions, and the presence of close
neighbors can contribute to the scatter in BT for S0, S,
and I galaxies. 
Moreover, the spread in
BT range is not surprising given that spheroids do not all have
$r^{1/4}$ profiles, and not all disks are pure exponentials 
\citep[e.g.,][]{graham03}.

\citet{simard08} derived the fraction of early-type galaxies, $f_e$,
using the same GIM2D criteria as in this Letter.  We use their
definition for early-types as being galaxies with BT$\ge$0.35 and
S2$\le$0.075 (c.f., \citet{tran01}).  The values of these selection
criteria are indicated by a vertical and horizontal line in
Fig.~\ref{fig1}. 

We find that many galaxies that qualify
as late-types by the quantitative criteria, but that are visually
classified as S0, generally have moderate quantitative asymmetry.

%%%%%%%%%%%%%%%%%%%%%%%%%%%%%%%%%%%%%%%%%%%%%%%%%%%%%%%%%%%%%%%%%%%%%%%%%%%
\section{Early-type Fractions}

The early-type fraction ($f_e$) depends both on how $f_e$ is defined
(visual vs. GIM2D) and the magnitude limit.  
The visual and quantitative morphologies yield nearly identical fractions.
We derive $f_e$ for
different magnitude limits and list the results for SG1120 in
Table~\ref{tbl-1}.\footnote 
{While mass selection is preferable to luminosity selection \citep[e.g.,][]{holden06,holden07}, 
we opt for luminosity
selection in this instance to facilitate comparison with existing samples. We note however that
for a stellar mass-limited sample we obtain high $f_e$ values comparable to cluster values from
\citet{holden06,holden07}.
} 
Generally,
we find $f_e$$\sim$70\% at a cutoff of $M^{\ast}$+0.5 and
$f_e$$\sim$60\% when including fainter members.

The latter value is
comparable to the fraction of passive galaxies in SG1120 with [OII]$\lambda$3727 
emission $<$5\AA~\citep[$\sim$61$\pm$8\%,][]{gonzalez05}.

\subsection{$f_e$ for Clusters}

Table~\ref{tbl-2} lists $f_e$ for galaxy clusters at a similar $z$ as
that of SG1120.  Independent of magnitude cutoff, clusters at these
redshifts have early-type fractions of $\sim$60\% \citep[within 
$\sim$500$h^{-1}_{65}$~kpc; see][and references therein]{lubin02} 
and also \citet{vandokkum01,holden04}.

We expect SG1120 to evolve into a cluster that is similar in
mass to \objectname{Coma}. We find that Coma and SG1120 have comparable early-type
fractions when similar magnitude limits are used: \citet{holden06}
measure $f_e$$\sim$78$\pm$4\% for luminosity-selected galaxies ($M^{\ast}$+0.5).

\subsection{$f_e$ for Groups and the Field at 0.3$<z<$0.55}

Table~\ref{tbl-2} also lists early-type fractions for galaxy groups in
the same redshift range.  The first two entries in the group section
refer to X-ray selected groups, and the remaining entries to
kinematically selected groups. 

\citet{jeltema07} provide data for
four X-ray luminous groups. \citet{mulchaey06} provide two additional X-ray luminous
groups. \citet{wilman05} report the fraction of
passive galaxies, as defined by [OII]$\lambda$3727, within 1$h^{-1}_{75}$~Mpc of the
centers of kinematically selected groups; we assume that the passive
fraction corresponds to $f_e$, as is the case for SG1120.

The scatter in $f_e$ is large among the galaxy group studies, but the
mean early-type fraction is distinctly higher than that of the field.

The field $f_e$ is 19$^{+6}_{-5}$\% at 0.2$\leq$$z$$\leq$
0.6 \citep{jeltema07}; comparable to that of the passive fraction in the field
\citep[$\sim$25\% for 0.2$\leq$$z$$\leq$0.6;][]{wilman05}.

\subsection{S0-to-E Ratio}

The ratio of S0 to elliptical galaxies in clusters is on average 0.5 at these 
redshifts \citep{dressler97,desai07}.  However, no measurement
of the S0/E ratio has been measured for the field or for galaxy groups
at intermediate redshifts. 

Using the visual classifications for all
confirmed members in SG1120, we measure the S0/E ratio to also be 0.5,  
similar to clusters at the same redshift.

\begin{figure}
\plotone{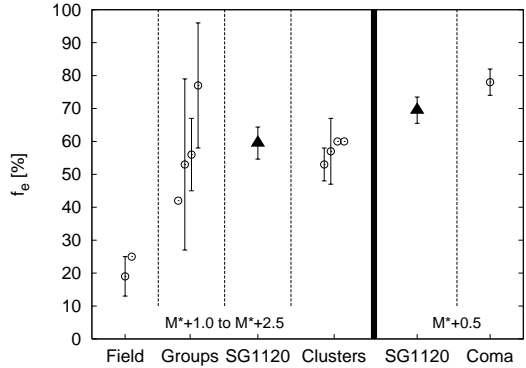}

\caption{The early-type fractions as described in the text for the
different environments, which are separated by dashed vertical
lines. SG1120 is shown as a solid triangle and literature values as
circles. SG1120 values are derived from the mean of the visual and GIM2D morphologies. 
The solid vertical line separates the galaxy groups and
clusters at 0.3$<$$z$$ \la$ 0.55 evaluated with magnitude cutoffs ranging from
$M^{\ast}$+1.0 to $M^{\ast}$+2.5 (left panel), from a comparison to Coma ($M^{\ast}$+0.5 right panel).  }\label{fig3}

\end{figure}

\begin{deluxetable}{llllll}
\tabletypesize{\scriptsize}
%\rotate
\tablecaption{Early-type fractions in SG1120 [\%]\label{tbl-1}}
\tablewidth{0pt}
\tablehead{
\colhead{} & \colhead{$M^{\ast}$+0.5} & \colhead{$M^{\ast}$+1} & \colhead{$M^{\ast}$+1.4} & \colhead{$M^{\ast}$+1.5} &
\colhead{$M^{\ast}$+1.75} }
\startdata
visual & 73$\pm$4 & 60$\pm$5 & 61$\pm$6 & 61$\pm$6 & 60$\pm$6 \\
GIM2D  & 66$\pm$4 & 57$\pm$4 & 59$\pm$4 & 60$\pm$4 & 58$\pm$4 \\
\enddata
\tablecomments{Table \ref{tbl-1} lists for the supergroup the $f_e$ in
percentages for both the visual and automated GIM2D classification;
$f_e$ is determined using multiple magnitude cut-offs because, $e.g.$
\citet{simard08} and \citet{poggianti06} use $M^{\ast}$+1.4 in their
sample.  The error for the visual results is the standard deviation
from the mean value of the visual classification from the five
classifiers.   }
\end{deluxetable}

%%%%%%%%%%%%%%%%%%%%%%%%%%%%%%%%%%%%%%%%%%%%%%%%%%%%%%%%%%%%%%%%%%%%%%%%%%%

\section{Results and Discussion}

Figure~\ref{fig3} presents $f_e$ as a function of environment.  On
the left side, the values for the field, groups, and clusters for the
magnitude cutoffs ranging from $M^{\ast}$+1.0 to $M^{\ast}$+2.5 are shown.  Comparing
the full samples is reasonable because we found that $f_e$ is nearly 
independent of the exact limit for cutoffs in this magnitude range. 
SG1120's early-type fraction is the mean of the
visual and quantitative GIM2D classification, and its mean error is from
Table~\ref{tbl-1}.  The right side of Fig.~\ref{fig3} shows $f_e$ for
SG1120 and Coma using the same magnitude limit ($M^{\ast}$+0.5).

The early-type fraction in the field is low and comparable to that
measured in the local Universe \citep[e.g.,][]{tran01}.  As shown in
Fig.~\ref{fig3}, $f_e$ for galaxy clusters at the redshift range of
SG1120 is $\sim$60\% on average
\citep{dressler97,vandokkum01,lubin02,holden04,desai07,simard08} and
thus similar to the $f_e$ measured for SG1120.  In addition, SG1120's
S0/E ratio is the same as that observed in galaxy clusters at the
redshift of SG1120.

The early-type fraction in the galaxy groups has a larger range
\citep{mulchaey06,jeltema07}; while the mean value falls below the
values of SG1120 and the galaxy clusters, this may not be the most
appropriate method for characterizing groups. For example,
\cite{poggianti06} has found that the range in passive galaxy fraction
corresponds to other physical characteristics, $e.g.$ velocity
dispersions of the groups.  Therefore, the group population likely
contains a range of systems that include pre-processed groups such as
those that make up SG1120, and less evolved galaxy groups.

The Coma cluster and SG1120 ($z$$\sim$0.37) 
both have almost comparable early-type
fractions when considering the errorbars ($M^{\ast}$+0.5).  Therefore, the
morphological mix in SG1120, a system made of four distinct X-ray
luminous galaxy groups that will assemble into a galaxy cluster, is
similar to that of clusters in the local universe.  We
conclude that galaxies in SG1120 are morphologically pre-processed in
the group environment and so the galaxy population as a whole does not
require additional morphological evolution.

While there must be some subsequent evolution due to processes
such as infall of field galaxies, mergers, and ram pressure, what
we learn from this system is that the net effect of these processes
need not be large. The $f_e$ values in SG1120 indicate that late-time infall
of groups does little to change the cluster early-type
fraction.  \citet{holden06,holden07} have previously demonstrated that 
there is little evolution in $f_e$ for massive cluster galaxies, which is to
be expected if late time accretion is dominated by groups like those in SG1120. 
If late time accretion is instead dominated by individual galaxies rather than groups,
as suggested by \citet{berrier08}, then our results imply that the accretion
and transformation rates (from local and global processes) must roughly balance to maintain a stable $f_e$.

Although SG1120 is only a single system, it
demonstrates that the cluster environment is not required to reach high
early type fractions, and that in general infalling groups should not significantly alter 
cluster $f_e$ values.
In fact, we see that in these group environments
mergers among even the most massive galaxies are common \citep{tran08}
supporting the hypothesis that mergers in the group environment are
a driver of galaxy evolution.

The large scatter in $f_e$ among galaxy groups, irrespective of
whether they are X-ray or kinematically selected, suggests that
obtaining a clean comparison sample of groups that will be accreted by
clusters is key to any evolutionary study. Thus while the weakness of
our study is having only 
a system, representing one possible cluster assembly scenario, 
it is mitigated by the knowledge
gained from studying galaxies in groups that are on the verge of
entering the cluster environment.

%%%%%%%%%%%%%%%%%%%%%%%%%%%%%%%%%%%%%%%%%%%%%%%%%%%%%%%%%%%%%%%%%%%%%%%%%%%

\acknowledgments

We thank the anonymous referee for a thoughtful, constructive report.
We are grateful to L. Simard for support with GIM2D. 
This work was supported by grant HST-GO-10499. 
KT acknowledges support from the Swiss National Science Foundation (grant
PP002-110576), and thanks J. Blakeslee for help during the initial ACS
reduction. JM acknowledges funding support from NASA-06-GALEX06-0030 
and Spitzer G05-AR-50443.

{\it Facilities:} \facility{HST (ACS), VLT (VIMOS, FORS2), Magellan (LDSS3), KPNO Mayall (FLAMINGOS)}.

\begin{deluxetable}{llllllll}
\tabletypesize{\scriptsize}
%\rotate
\tablecaption{Early-type fractions of Groups and Clusters\label{tbl-2}}
\tablewidth{0pt}
\tablehead{
%\colhead{} & \colhead{$f_e$~[\%]} & \colhead{$z$} & \colhead{cut~[$M^{\ast}$+]} & \colhead{$\sigma$~[km~s$^{-1}$]}&\colhead{\#} &\colhead{method}  }
\colhead{} & \colhead{$f_e$} & \colhead{$z$} & \colhead{cut} & \colhead{$\sigma$}&\colhead{\#} &\colhead{method}  \\
\colhead{} & \colhead{[\%]} & \colhead{} & \colhead{[$M^{\ast}$+]} & \colhead{[km~s$^{-1}$]}&\colhead{} &\colhead{}  }
\startdata
{\bf Clusters:}&&&&&&\\
\citet{simard08}   & 53$\pm$5 & 0.3-0.55& 1.4  & 681-1080 & 4 & GIM2D\\
\citet{dressler97}  &57$\pm$10 & 0.3-0.55& 2.5 & n.s. & 10 & visual\\
\citet{desai07}\tablenotemark{a}\tablenotemark{b}     & $\sim$60&$\sim$0.4& 1.5 & $>$600 & 12 & visual\\
\citet{lubin02}\tablenotemark{b}     & $\sim$60&$\sim$0.4& 1.5  & n.s. & 11 & visual\\
\hline
{\bf Groups:}&&&&&&\\
\citet{jeltema07}\tablenotemark{c}   &77$\pm$19 &0.3-0.50  & 1.4 &211-417 & 4 & visual\\
\citet{mulchaey06}  & 53$\pm$26&0.3-0.50  & 1.0  & 245-632& 2 & visual\\
\citet{simard08}    & 56$\pm$11&0.3-0.55 & 1.4 & 165-540 & 7 & GIM2D\\
\citet{wilman05}    & 42&0.3-0.55 & 1.75  & 100-800 & 26 & [OII]\\
\enddata
\tablecomments{Table \ref{tbl-2} contains the $f_e$ of clusters and
groups from in the literature.  Col.~3 shows the magnitude cutoffs
that were used to derive $f_e$.  The range in velocity dispersion for
the systems are given in Col.~4; if not specified in the literature,
we use the `n.s.' space holder.  Col.~5 contains the number of
clusters/groups studied by each reference.  The classification method
(GIM2D, visual, or [OII]$\lambda$3727 emission) is noted in Col.~6.}
\tablenotetext{a}{The galaxy cluster sample in this study overlaps 
with the sample in \citet{simard08}}
\tablenotetext{b}{The galaxy clusters studied in these references
partially overlap with the systems of \citet{dressler97}. }
\tablenotetext{c}{Here we recomputed the $f_e$ values from the provided galaxy table for different 
magnitude cutoffs and found that $f_e$ is constant over the range between $M^{\ast}$+0.5 and
$M^{\ast}$+2.5. We therefore show only one value.}
\end{deluxetable}

\end{document}